\documentclass[useAMS,usenatbib]{mn2e}
\usepackage[dvips]{graphicx}

\title[A study of methyl formate in astrochemical environments]
  {A study of methyl formate in astrochemical environments}
\author[A. Occhiogrosso et al]
  {A.~Occhiogrosso,$^1$\thanks{E-mail: ao@star.ucl.ac.uk.}
  S. Viti,$^1$ P. Modica,$^2$ M. E. Palumbo,$^2$ \\
  $^1$Dept. of Physics and Astronomy UCL, Gower Place, London WC1E6BT, UK \\
  $^2$INAF-Osservatorio Astrofisico di Catania, via Santa Sofia 78, 95123, Catania, Italy}
\date{Released 2011 Xxxxx XX}

\pagerange{\pageref{firstpage}--\pageref{lastpage}} \pubyear{2011}

\def\LaTeX{L\kern-.36em\raise.3ex\hbox{a}\kern-.15em
    T\kern-.1667em\lower.7ex\hbox{E}\kern-.125emX}

\begin{document}

\maketitle

\begin{abstract}

Several complex organic molecules are routinely detected in high
abundances towards hot cores and hot corinos. For many of them, their
paths of formation in space are uncertain, as gas phase reactions alone
seem to be insufficient.

In this paper, we investigate a possible solid-phase 
route of formation for methyl formate (HCOOCH$_{3}$). 
We use a chemical model updated with recent results from an experiment where simulated grain surfaces
were irradiated with 200 keV protons at 16 K, to simulate the effects of cosmic ray irradiation on
grain surfaces. 

We find that this model may be sufficient to reproduce the observed methyl formate 
in dark clouds, but not that found in hot cores and corinos.

\end{abstract}


\section{Introduction}

Methyl formate (HCOOCH$_{3}$) is the simplest example of an ester. It
is derived from the formic acid (HCOOH), where a methyl group is
attached to the carboxyl group. It is an important organic complex
molecule that was first detected by Brown et al. (1975) towards the Sgr
B2(N), the richest molecular source in the galaxy located in the
Galactic Centre giant cloud Sgr B2. It is considered to play a key
role in understanding the origin of life because it leads to the
synthesis of bio-polymers. Methyl formate has two structural isomers,
glycolaldehyde (HCOCH$_{2}$OH) and acetic acid (CH$_{3}$COOH), but it
has been reported that this molecule is the most abundant among these
isomers (Hollis et al. 2001). Particularly, its column density in the
Orion Hot Core was derived to be 9.4 x 10$^{15}$ cm$^{-2}$ by
Ikeda et al. (2001) and this value was confirmed by
Sakai et al. (2007). Methyl formate has also been detected in the
G31.41+0.31 hot molecular core (HMC); its column density is observed
to be 3.4 x 10$^{18}$ cm$^{-2}$ around this region. Cazaux et al. (2003)
observed methyl formate in the hot core around the protostellar object IRAS 16293-2422.
Hence methyl formate seems to be ubiquitous in star forming regions.\\ 
\indent Several studies have been carried out
on methyl formate to understand its formation mechanisms, but it is
still debated whether complex organic molecules form on dust icy
mantles during the cold phase, on grains during the warm-up phase or in the gas phase. 
Despite a large activation energy barrier
between protonated methanol and formaldehyde (Horn et al. 2004),
Garrod et al. (2006) found that the latter route of formation is viable during
the icy mantle sublimation phase. Moreover, 
very recently, Laas et al. (2011) have
proposed two Fischer esterification (the acid-catalysed reaction of a
carboxylic acid with an alcohol to give an ester) pathways that occur
during the warming-up phase, involving protonated formic acid and
methanol and protonated methanol and neutral formic acid,
respectively. Both reactions have two channels that correlate to cis-
trans- protonated methyl formate. They emphasised that methanol
photodissociation branching ratios and warm-up timescales influence
the relative ratios between these two geometries.

The possibility that more complex species are formed at low temperatures on surfaces has been investigated before (e.g. Charnley 1997, 2001; Herbst \& van Dishoeck 2009)
and in particular methyl formate production on dust surfaces was first proposed by Herbst (2005): CO, C and O lead to the formation of CH$_{3}$O and
HCO radicals, both known as methyl formate precursors. Based on this
pathway and using energetic electrons at 10 K, laboratory experiments
were performed by Bennet \& Kaiser (2007) in order to produce methyl formate
and estimate the rate coefficients for this reaction. Different
experiments involving methyl formate were also performed by
Gerakines et al. (1996) and Oberg et al. (2009), which obtained this molecule
after UV photolysis of pure methanol and CO:CH$_{3}$OH ice
mixtures. Recently, a laboratory study by Modica \& Palumbo (2010) has
suggested a new solid state route of formation for this molecule. By
using infrared spectroscopy in the 4400 - 400 cm$^{-1}$ range for in
situ monitoring the sample during the experiments, they simulated a
cosmic ion irradiation on a binary mixture containing CH$_{3}$OH and CO
ice: methyl formate was therefore released to the gas-phase after
sublimation of these icy samples.

In the present work, we include the Modica \& Palumbo (2010) experiments in
our chemical model UCL\_CHEM Viti et al. (2004a), by extrapolating from the
experiments a new rate coefficient for the methyl formate formation on
grains.  Our purpose is to investigate whether surface reactions
during the cold phase $alone$ can account for the observed abundances of
methyl formate in massive star-forming regions and around low-mass
stars. In Section 2 we describe the experimental procedure; physical
and chemical details of the chemical model are summarised in Section
3. In Section 4 we qualitatively model four different sources, and present our findings.

\section{Experiments}

Experiments have been performed in the Laboratory of Experimental Astrophysics in Catania (Italy).
Solid samples were prepared and irradiated in a stainless steel vacuum chamber where pressure is kept below 10$^{-7}$ mbar. The gas mixture to be investigated was injected into the chamber through a needle valve where it  froze onto the substrate (Si or KBr) placed in thermal contact with the tail section of a cryostat (10-300 K).
After deposition the samples were bombarded by 200 keV H$^+$ ions. Ions are obtained from an ion implanter interfaced with the vacuum chamber. The beam used produces current densities in the range from 0.1 to a few $\mu$A cm$^{-2}$ in order to avoid macroscopic heating of the target.
Infrared transmittance spectra of the samples were obtained  before and after each step of irradiation by a FTIR spectrometer (4400-400 cm$^{-1}$=2.27-25 $\mu$m).
More details on the experimental procedure can be found in Palumbo et al. (2004) and in Modica \& Palumbo (2010).

Pure CH$_3$OH and a CO:CH$_3$OH mixture were irradiated at 16 K with 200 keV H$^+$ ions.
In both cases, after irradiation, the IR spectra show several absorption bands which testify for the formation of new molecules not present in the original sample. Among these, two bands due to CH$_4$ appear at 3010 cm$^{-1}$ and 1304 cm$^{-1}$ (Palumbo et al. 1999);  CO$_2$ bands appear at 2344 and 660 cm$^{-1}$ ($^{12}$CO$_2$) and 2278 cm$^{-1}$ ($^{13}$CO$_2$); H$_2$CO is detected at 1720 cm$^{-1}$ (Hudson \& Moore 2000);  C$_2$H$_4$(OH)$_2$ (ethylene glycol) is observed at 1090 cm$^{-1}$ (Hudson \& Moore 2000). Moreover  a band is observed near 1160 cm$^{-1}$ and is attributed to methyl formate (HCOOCH$_3$) and a band near 1067 cm$^{-1}$ is assigned to glycolaldehyde (HCOCH$_2$OH) (Modica \& Palumbo 2010).

Figure 1 shows the ratio between the column density of methyl formate and methanol, N(HCOOCH$_3$)/N(CH$_3$OH),  as a function of ion fluence (ions cm$^{-2}$), measured after each step of irradiation.
The experimental data have been fitted with an exponential curve
\begin{center}
N(HCOOCH$_3$)/N(CH$_3$OH)=A(1-e$^{- \sigma \Phi}$)
\end{center}

where A is the asymptotic value,  $\sigma$ is the cross section (cm$^2$) and $\Phi$ is the ion fluence (ions cm$^{-2}$).

The fitting procedure gives A=0.0042 and $\sigma$=8.6$\times$10$^{-15}$ cm$^2$.

\begin{figure}
\includegraphics[width=90mm]{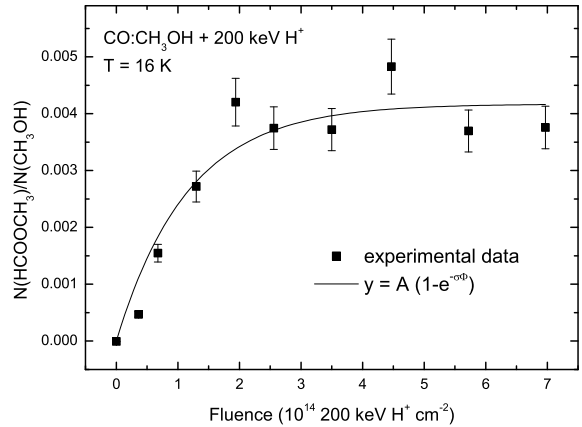}
\caption{The ratio between the column density of methyl formate and methanol, N(HCOOCH$_{3}$)/N(CH$_{3}$OH), as a function of ion fluence after irradiation of a CO:CH$_{3}$OH mixture at 16 K.}
\end{figure}

To apply the laboratory results to the interstellar medium conditions we assume a standard ionisation rate ($\zeta$ = 3$\times$10$^{-17}$ s$^{-1}$) and  derive the flux F$_{ISM}$=0.5 ions cm$^{-2}$ s$^{-1}$ of cosmic ions in the approximation of effective monoenergetic 1 MeV protons (see Mennella et al. (2003)).  F$_{ISM}$ must be regarded as an effective quantity. It represents the equivalent flux of 1 MeV protons, which gives rise to the ionisation rate produced by the cosmic-ray spectrum if 1 MeV protons were the only source for ionisation.
Furthermore we assume that the cross section scales with the stopping power (S, energy loss per unit path length) of impinging ions.
According to SRIM code (Ziegler et al. 2008), in the case of protons impinging on a CO:CH$_3$OH mixture S(200keV protons)=2.9$\times$S(1MeV protons).

We defined $\sigma_{ISM}$= A$\times \sigma$/2.9 cm$^2$, where the product A$\times \sigma$ is the initial slope of eq. (1).
The formation rate of methyl formate is given by 

R(s$^{-1}$)= $\sigma_{ISM}$(cm$^2$) $\times$ F$_{ISM}$(cm$^{-2}$ s$^{-1}$) 

Using $\sigma_{ISM}$ and F$_{ISM}$ as estimated above we obtain R=6.2$\times$10$^{-18}$ s$^{-1}$.

\section{Chemical model}

The chemical model is UCL\_CHEM, developed by Viti \& Williams (1999) and
Viti et al. (2004a).  UCL\_CHEM is a time and depth dependent gas-grain
chemical model which estimates the fractional abundances (with respect
to the total number of hydrogen nuclei) of gas and surface species in
every environment where molecules are present. For this purpose it was
adapted to model low and high-mass star forming regions.  The model
performs a two-step calculation: Phase I starts from a fairly diffuse
medium in atomic form and undergoes a free-fall collapse until
densities typical of the gas that will form hot cores and hot corinos are reached (10$^{7}$ cm$^{-3}$  
and 10$^{8}$ cm$^{-3}$ respectively). During this time atoms and molecules are
depleted onto grain surfaces and they hydrogenate when possible. The
depletion efficiency is determined by the fraction of the gas-phase
material that is frozen onto the grains. This approach allows a
derivation of the ice composition by a time-dependent computation of
the chemical evolution of the gas-dust interaction process. The
initial elemental abundances of the main species (such as H, He, C, O,
N, Si and Mg) are the main input for the chemistry. In our model surface reactions only occur during this phase.
Phase II is the warming-up phase and follows the chemical evolution of the remnant
core when the hot core itself is formed. It simulates the effect of
the presence of an infrared source in the centre of the core or in its
vicinity by subjecting the core to an increase in the gas and dust
temperature. We derived the temperature of the gas as a function of
the luminosity (and therefore the age) of the protostar. The treatment
of the evaporation of the code is either time-dependent in which
mantle species desorb in various temperature bands according to the
experimental results by Collings et al. (2004) or instantaneous in that all
species will desorb off grain surfaces in the first time-step. The
temperature profile is another input parameter. For this study, the model 
does not take into consideration the desorption of molecules due to UV photons and cosmic rays.

Most of the gas-phase reactions that occur in the ISM are collected
from the UMIST database (Woodall et al. 2007). Because of the failure of
gas-phase chemistry to reproduce abundances of particular molecules,
in the present work a gas-grain network reaction is also taken into
account and particularly, we focused on methyl formate reactions.
Table 1 shows on the top the updated gas-phase reaction rate
coefficients as listed in the most recent database KIDA
(Wakelam 2009). At the bottom, we report solid state paths of
formation of methyl formate, taking into account the new experimental
rate coefficient for this reaction.  Methyl formate is both one of the
reactants (R) and product (P); $\alpha$, $\beta$, and $\gamma$ are the
rate coefficients. The first six are bimolecular reactions, where an
exothermic proton transfer occurs between an ion and a neutral
species. The last one is an electronic recombination, in which the
recombination of a positive ion and an electron results in the
dissociation of the molecule. In both cases, a Koiij formula (a
modified version of the Arrhenius equation) is adopted to evaluate the
reaction rate:

\begin{equation}
\rmn{k}(T) = \alpha(\frac{T}{300})^{\beta}e^{-\frac{\gamma}{T}},
\end{equation}

where T is the gas temperature, $\alpha$ is the cosmic ion rate,
$\beta$ contains the dipole effect and $\gamma$ represents the
probability per cosmic ray ionisation, that for the updated methyl
formate reactions is always zero. $\beta$ includes the enhancement of
ion-neutral rate coefficients for cases in which the neutral has a
large, permanent electric dipole moment. This effect results in a rate
coefficient that has a T$^{1\over 2}$ dependence at low temperature
(10 - 50 K) as observed by Herbst \& Leung (1986). In the bottom panel of
Table 1 the surface path of formation of methyl formate is summarised:
one of the most abundant species in the ISM is carbon monoxide, that
undergoes successive hydrogenations in the presence of a large amount
of hydrogen (Fuchs et al. 2009; Watanabe et al. 2007). Consequently, methanol can be
formed. After cosmic ion irradiation, CO reacts with CH$_{3}$OH to
produce methyl formate (see Modica \& Palumbo (2010)). The $m$ before the
molecular formul\ae \ has been inserted to underline the role of the
surface chemistry that acts as a catalyst for the reactions with large
energy barriers. The new experimental rate coefficient for this
reaction is 6.2 x 10$^{-18}$ s$^{-1}$, as shown.

\begin{table*}
 \begin{minipage}{126mm}
  \caption{Gas-phase and solid state routes of methyl formate destruction and formation. 
The top panel shows the gas-phase reactions involving  methyl formate, 
whose rate coefficients are taken from the KIDA database. 
The bottom panel contains the new experimental path of methyl formate formation 
on grain surface as investigated by Modica \& Palumbo (2010).}
  \label{symbol}
  \begin{tabular}{@{}|cccccccc}
  \hline
R1 & R2 & P1 & P2 & P3 & $\alpha$ & $\beta$ & $\gamma$\\
   \hline
HCOOCH$_{3}$ & H$_{3}^{+}$ & H$_{2}$ & H$_{5}$C$_{2}$O$_{2}^{+}$ & & 4.05E-09 &- 0.5 & 0\\
HCOOCH$_{3}$ & He$^{+}$ & He & CH$_{3}$ & HCO$_{2}^{+}$ & 3.54E-09 & -0.5 & 0\\
HCOOCH$_{3}$ & H$_{3}$O$^{+}$ & H$_{2}$O & H$_{5}$C$_{2}$O$_{2}^{+}$ & & 1.81E-09 & -0.5 & 0\\
HCOOCH$_{3}$ & HCO$^{+}$ & CO & H$_{5}$C$_{2}$O$_{2}^{+}$ & & 1.55E-09 & -0.5 & 0\\
HCOOCH$_{3}$ & C$^{+}$ & COOCH$_{4}^{+}$ & C & & 2.17E-09 & -0.5 & 0\\
HCOOCH$_{3}$ & H$^{+}$ & COOCH$_{4}^{+}$ & H & & 6.9E-09 & -0.5 & 0\\
H$_{5}$C$_{2}$O$_{2}^{+}$ & e$^{-}$ & HCOOCH$_{3}$ & H & & 1.5E-07 & -0.5 & 0\\
   \hline
mCH$_{3}$OH & mCO & mHCOOCH$_{3}$ & & & 6.20E-18 & 0 & 0\\
   \hline
  \end{tabular}
 \end{minipage}
\end{table*}
   
\section{Results}

Figure 2 shows the fractional abundances of selected species (from
Table 1) as a function of time during Phase I. The solid lines
represent the abundances of methanol, carbon monoxide and methyl
formate when the new solid state route is included in the model. The
dashed line is methyl formate obtained from the model computed with the same chemistry as in 
the original code (Viti et al. 2004a). In the latter case, the reactant
trends are the same as those shown in the updated version of the
code. While the reactants, CO and CH$_{3}$OH, reach a fractional abundance of 4 x 10$^{-5}$ and 2 x 10$^{-6}$
respectively, methyl formate is produced in a negligible quantity ($<$
10$^{-21}$) onto the grains. By considering the solid state path of methyl formate
formation with the experimental rate coefficient of 6.2 x 10$^{-18}$
s$^{-1}$ (as described in Sections 2 and 3) a value of 1.8 x 10$^{-9}$
is reached for its fractional abundance (solid line). 
In order to compare our theoretical abundances with those from the observations, we also run Phase II for a typical hot core.
In order to facilitate the calculation we set an instantaneous evaporation instead of a time-dependent sublimation for molecules that sublimate from the grain into the gas phase.
As a result we obtained a detectable abundance of methyl formate with a value of 1.7 x 10$^{-9}$. 
An alternative route for the synthesis of methyl formate in a hot core
was previously investigated by Garrod \& Herbst (2006), that proposed both
gas-phase and grain-surface processes to produce most species during
the warm-up phase. They estimated that the gas-phase/accretion path,
involving protonated methanol and formaldehyde, is responsible for the
formation of 25\% of the total HCOOCH$_{3}$ present on the grain
surface before sublimation. Indeed, during the warm-up phase, the
surface reaction between CH$_{3}$O and HCO radicals is allowed and it
is thought to be the dominant mechanism; when significant amount of
H$_{2}$CO begins to evaporate, the methyl formate formation on grains
is inhibited and formaldehyde is protonated to produce methyl formate
on gas-phase. These reactions are more efficient in hot corinos
than in hot cores. They found a trend where methyl formate reaches a fractional abundance of about
10$^{-8}$ at 10$^{6}$ yr. Both mechanisms seem therefore to be viable.

In order to compare
the theoretical abundance of methyl formate with the observational data,
we model two high massive forming regions, the Orion Hot Core and the
G31.41+0.31, a low-mass source, NGC1333- IRAS2 and the cold gas near the region B1-b in Perseus using our two-phases UCL\_CHEM. The physical and chemical parameters for each model are
listed below. In Phase II of all models we now assume a time-step
sublimation instead of an instantaneous evaporation for molecules that
sublimate from the grains into the gas-phase during the warm-up stage
of star formation process, as described by Viti et al. (2004a).

\begin{figure}
 \includegraphics[width=60mm, angle=270]{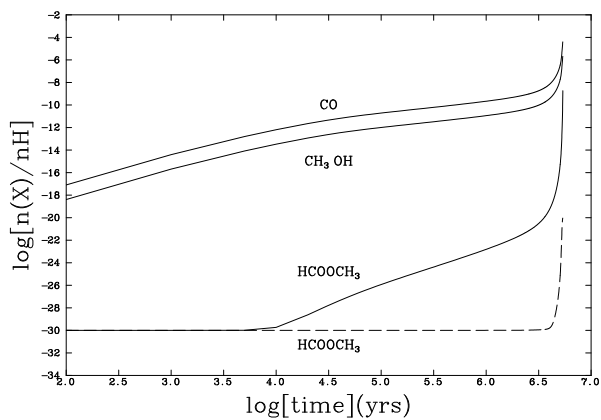}
 \caption{Fractional abundances (with respect to the total number of hydrogen nuclei) of selected species as a function of time.}
\end{figure}

\begin{table*}
 \begin{minipage}{126mm}
  \caption{Observed and predicted column densities towards Orion KL Hot Core, G31.41+0.31, NGC1333-IRAS2 and B1-b core.}
  \label{anymode}
  \begin{tabular}{@{}|cccccc}
  \hline
Source & Type & Distance(pc) & T$_{rot}$(K) & Observations(cm$^{-2}$) & Model(cm$^{-2}$)\\
   \hline
Orion KL Hot Core & High mass & 480 & 250 & $^{a}$6.9x10$^{16}$ & 3.5x10$^{14}$\\
G31.41+0.31 HMC & High mass & 7900 & 300 & $^{b}$6.8x10$^{18}$ & 9.3x10$^{15}$\\
NCG1333-IRAS2 & Low mass & 220 & 38 & $^{c}$5.8x10$^{14}$ & 6.4x10$^{12}$\\
B1-b & Dark core & 350 & $<$30 & $^{d}$8.3x10$^{12}$ & 1.2x10$^{12}$\\
   \hline
  \end{tabular}
\footnotetext[1]{Remijan et al. (2003)}
\footnotetext[2]{Note that the total methyl formate abundance reported herein is twice the value of form-E given in Cesaroni et al. (1994)}
\footnotetext[3]{Note that the total methyl formate abundance reported herein is twice the value of form-A given in Bottinelli et al. (2007)}
\footnotetext[4]{Oberg et al. (2010)}
 \end{minipage}
\end{table*}

\subsection*{Orion KL Hot Core}

The Orion Molecular Cloud 1 (OMC1) is a dense clump at 480 pc that
contains several distinct infrared-emitting regions. Among these the
Kleinmann-Low (KL) region is composed of four different components:
the hot core, the compact ridge, the plateau and the photodissociation
region (PDR) surrounding the quiescent gas. While the hot core and
the plateau are characterised by elevated temperatures, the ridge
consists of an extended and cooler region with quiescent material. We
focused on the Orion hot core as it is one of the richest known
astronomical sources of molecular lines. The main heating mechanism in
this location is radiative, and gas kinetic temperatures are close to
200 K. Using UCL\_CHEM Lerate et al. (2008) computed different chemical
models to simulate each component of the KL region. In the present
work we used the physical and chemical parameters for their best-fitting
model. 

A comparison between our model calculations and the observational
column densities through the Orion Hot Core is shown in Table
2. Observational data are taken from Remijan et al. (2003), that have
surveyed this source with the BIMA Array with angular resolution of
about 2$^{\prime\prime}$-5$^{\prime\prime}$.

\subsection*{G31.41+0.31}

G31.41+0.31 is another example of a multiple massive star formation
region located at 7.9 kpc. It is associated with an hot molecular core
(HMC) of 0.048 pc in size, where the temperature is found to be about 300 K and the
luminosity is about 3 x $10^{5}$ solar luminosities
(Churchwell et al. 1990). It is in fact the region where the simplest
sugar, glycolaldehyde (and also one of the methyl formate isomers) was
first detected by Beltran et al. (2009) through IRAM PdBI
observations. Beltran et al. (2009 also used UCL\_CHEM to try to
reproduce the abundance for glycolaldehyde. We adopted the physical
and chemical input parameters from their study for our model.

We compared our theoretical results with those given by
Cesaroni et al. (1994), that performed high resolution observations by
using IRAM Plateau de Bure Interferometer. 

\subsection*{NGC 1333-IRAS2}

Located in the cloud NGC1333, belonging to the Perseus complex at 220
pc in distance, IRAS2 is a low-mass binary protostellar system,
including IRAS2A and IRAS2B, separated by $30''$. An infalling
envelope, a circumstellar disk and multiple outflows seem to be
associated with them.  The advantage of observing such nearby objects is that
these sources can be spatially resolved with millimetre
interferometers. As a result we adopted a multipoint model in order to
reproduce the density profile of the hot corino associated with IRAS2, as reported in
Maret et al. (2004). We also used the temperature profile for
hot corino given in Awad et al. (2010).

We evaluated the total column density for the methyl formate as
described in Viti et al. (2004b):

\begin{equation}
N(HCOOCH_{3}) = \sum(X \times A_{\rm v} \times L) \times N(H_{2}),
\end{equation}

where X is the methyl formate fractional abundances and $A_{\rm v}$ is the visual extinction at length L (pc).

\subsection*{B1-b core}

The protostar associated with the B1-b core is believed to be between a pre-stellar and Class 0 protostar evolutionary stage. Two sources are identified with the B1-b core, designed as B1-bN and B1-bS, separated by $20''$ in the north-south direction at about 350 pc in distance (Hirano 1999). The physical parameters of two sources are very similar (T$_{dust}$ 18 K, mass of 1.6-1.8 solar masses and luminosity about 2.6-3.1 solar luminosities). We qualitatively model the B1-b core running our two-phases model; for this core, during Phase II the temperature only rises up to 30 K; this implies that only the weakly bound species partly sublimate due to thermal desorption (see Collings et al. (2004)); hence any methyl formate formed on the grains would remain in solid phase. However, non thermal desorption processes are known to be efficient even at 10 K (see Roberts et al. (2007); Boland \& de Jong (1982); Hasegawa \& Herbst (1993) and Bringa \& Johnson (2004)). We have therefore ran Phase I including some non thermal desorption mechanisms (as in Roberts et al. (2007)). We compare our results for methyl formate with those reported in Oberg et al. (2010). Note that our computed methyl formate is a lower limit as our non thermal desorption mechanisms did not include direct UV photodesorption.

\indent Table 2 reports all our model results and compares the theoretical predicted column densities for methyl formate with those observed in the sources described above. 
A relatively good agreement within a factor of 10 was found for the 
dark core associated with B1, 
but the theoretical values are too low for the cases of warmer star forming regions.
In fact around protostars the flux of ions due to stellar flares should also be considered. As discussed by Garozzo et al. (2011), the effects induced by cosmic rays on icy grain mantles during the collapse phase era are comparable to the effects induced by stellar flare ions during the warm-up phase.

In summary, using the rate coefficient derived from laboratory
experiments of ice irradiation, we investigated the viability of a
cold solid state path to form methyl formate during the evolution of
protostellar cores when the temperatures are low (10 K). We find that
we cannot reproduce the abundance of methyl formate found
in hot cores and hot corinos without invoking gas- or
solid-phase reactions which necessitate high temperatures (and that
can therefore only occur during the warm up phase once the star is
formed), while we were able to reproduce the observed abundances in dark clouds. 
This work supports the idea that cosmic ion irradiation of
icy grain mantles may be able to contribute to the production of the
methyl formate observed in dense molecular clouds, but that other routes of formation are also required.

\section{Acknowledgments}
The research leading to these results has received funding from the [European Community's] Seventh Framework Programme [FP7/2007-2013] under grant agreement n$^{\circ}$ 238258.
We thank an anonymous referee for helping us improve the original manuscript.

\end{document}